\documentclass[a4paper,11pt]{article}

\usepackage{xspace}
\usepackage[utf8]{inputenc}
\usepackage{natbib}
\usepackage{amsmath,amssymb}

\newcommand\proglang[1]{\texttt{#1}\xspace}
\newcommand\pkg[1]{\texttt{#1}\xspace}
\newcommand\code[1]{\texttt{#1}\xspace}
\newcommand\sbm{\emph{stochastic block model}\xspace}
\newcommand\nR{\mathbf{R}}
\newcommand\nC[1]{\mathcal{C}_{#1}}

\usepackage{url}
\usepackage{gnuplot-lua-tikz}

\bibpunct{(}{)}{;}{a}{,}{,}

\title{Wmixnet: Software for Clustering the Nodes of Binary and Valued Graphs using the Stochastic Block Model}
\author{Leger, J.-B.
    \footnote{INRA, UMR 518 MIA, 16 rue Claude Bernard, Paris}
    \footnote{AgroParisTech, UMR 518 MIA, 16 rue Claude Bernard, Paris}
}
\date{}

\begin{document}

\maketitle

\section*{Abstract}
  Clustering the nodes of a graph allows the analysis of the topology of a
network.

  The \sbm is a clustering method based on a probabilistic
  model. Initially developed for binary networks it has recently been extended
to
  valued networks possibly with covariates on the edges.

  We present an implementation of a variational EM algorithm. It is written
  using \proglang{C++}, parallelized, available under a GNU General Public License (version
  3), and can select the optimal number of clusters using the ICL criteria. It
  allows us to analyze networks with  ten thousand nodes in a reasonable amount
  of time.

\section{Introduction}
  Complex networks are being more and more studied in different domains such as social
  sciences and biology. The network representation of the data is graphically
  attractive, but there is clearly a need for a synthetic model, giving an
  enlightening representation of complex networks. Statistical methods have been
  developed for analyzing complex data such as networks in a way that could
  reveal underlying data patterns through some form of classification.

  Unsupervised classification of the vertices of networks is a rapidly
  developing area with many applications in social and biological sciences. The
  underlying idea is that common connectivity behavior shared by several
  vertices leads to  their grouping in one {\it meta-vertex}, without losing too
  much information. Thus, the initial complex network can be reduced to a
  simpler {\it meta-network}, with few {\it meta-vertices} connected by few {\it
  meta-edges}.  \citet{PMDCR} show applications of this idea to biological
 networks and \citet{NS} and \citet{HRT} to social networks.

  Model-based clustering methods model the heterogeneity between nodes by
  grouping the nodes into classes. The model used in this paper is an extension of
  the \sbm (SBM) \citep{NS}. This model assumes that the
  nodes are distributed into groups, and connectivity between nodes is driven by
  node group memberships.

  SBM for non-binary graphs, with or without covariates has been introduced in
  \citet{Mariada10}. In this paper, a \emph{variational Expectation--Maximization}
  algorithm has been used to estimate parameters and to predict groups.

  This article introduces \pkg{wmixnet}, an implementation of the \emph{variational
  ex\-pectation--ma\-xi\-mi\-za\-tion} algorithm for this extension of the \sbm
  with or without covariates for three families of laws of probability:
  Bernoulli, Poisson, Gaussian.

  This implementation allows us to estimate parameters and to predict node groups
  and covariate effects for graphs which are valued or binary, directed or not,
  and with or without covariates.

\section{SBM model with covariates}

  We introduce here the \sbm with covariates and three
  probability distributions.

  \subsection{Notations}
    \paragraph{Graph.}
      Consider a graph $G=(V,E,w)$, where
      \begin{itemize}
        \item $V$ is the set of nodes, labelled in $\{1,\cdots,n\}$,
        \item $E$ is the set of edges, which is a subset of $V^2$,

        \item $w\colon E\rightarrow \nR$, is the function which gives
              edge weights.
        \item $Y\colon V^2\rightarrow \nR^p$, is the function which gives
              the covariate vector associated to each couple of nodes.
      \end{itemize}
      We  assume \emph{without loss of generality} that $E=V^2$, with the
      convention $w(i,j)=0$ if there is no edge from vertex $i$ to vertex $j$.

    \paragraph{Groups.}
      Consider $Q$ classes of nodes. For a given partition
      $\left(\nC{1},\cdots,\nC{Q}\right)$ of $V$, for a node $i$ and a group
      $q$, let $Z$ be defined as $Z_{iq}=1 \Leftrightarrow i\in\nC{q}$. And let
      $Z_i = \left(Z_{i1},\cdots Z_{iQ}\right)$.

  \subsection{The model}

    \paragraph{Nodes.}
      The class memberships of the nodes are driven by  independent
      identically distributed multinomial distributions:
      \[
        \forall i\in V \; Z_i \overset{\textrm{i.i.d.}}{\sim}
        \mathcal{M}(1,\alpha)
      \]
      where $\alpha=\left(\alpha_1,\cdots,\alpha_Q\right)$ and $\sum_q \alpha_q=1$.

    \paragraph{Edges.}
      For each couple of nodes $(i,j)$ the probability law of the link is driven
      by their class memberships and the $(i,j)$ covariate $Y(i,j)$:

      \[
        \left( w(i,j)|(i,j)\in\nC{q}\times\nC{l} \right) \sim
        \mathcal{F}_{ql}(Y(i,j)).
      \]

  \subsection{Probability laws}

      Generally, various probability laws can be used.  The probability
      distributions which are implemented in \pkg{wmixnet} are the following:
      \begin{itemize}
        \item \emph{Bernoulli}:
          \subitem without covariates:
                   $\mathcal{F}_{ql}(Y(i,j)) = \mathcal{B}(\pi_{ql})$. This
                   model does not use covariates and can model only binary
                   networks. This is the classical \sbm model.
          \subitem with covariates (with homogeneous effects):
                   $\mathcal{F}_{ql}(Y(i,j)) =
                   \mathcal{B}(\pi_{ql}\frac{1}{1+\exp(-\beta^TY_{ij}})$. This
                   model uses covariates and can model only binary
                   networks. The effect of covariates is the same for all pairs of
                   classes.
          \subitem with covariates (with heterogeneous effects):
                   $\mathcal{F}_{ql}(Y(i,j)) =
                   \mathcal{B}(\pi_{ql}\frac{1}{1+\exp(-\beta_{ql}^TY_{ij}})$.
                   This model uses covariates and can model only binary
                   networks. The effect of covariates is \emph{not} the same for
                   all pairs of classes.

        \item \emph{Poisson}:
          \subitem without covariates:
                   $\mathcal{F}_{ql}(Y(i,j))=\mathcal{P}(\lambda_{ql})$. This
                   model does not use covariates and can model networks with
                   non negative integer weights.
          \subitem with covariates (with homogeneous effects):
                   $\mathcal{F}_{ql}(Y(i,j)) = \mathcal{P}(\lambda_{ql}(Y(i,j))$
                   where $\lambda_{ql}(Y(i,j)=\lambda_{ql}\exp(\beta^TY(i,j))$.
                   This model uses covariates and can model networks with non
                   negative integer weight. The effect of covariates is the same
                   for all pairs of classes.
          \subitem with covariates (with heterogeneous effects):
                   $\mathcal{F}_{ql}(Y(i,j)) = \mathcal{P}(\lambda_{ql}(Y(i,j))$
                   where
                   $\lambda_{ql}(Y(i,j)=\lambda_{ql}\exp(\beta_{ql}^TY(i,j))$.
                   This model uses covariates and can model networks with non
                   negative integer weight. The effect of covariates is \emph{not}
                   the same for all pairs of classes.

        \item \emph{Gaussian}:
          \subitem without covariates:
                   $\mathcal{F}_{ql}(Y(i,j))=\mathcal{N}(\mu_{ql},\sigma^2)$.
                   This model does not use covariates and can model networks
                   with real weight.
          \subitem with covariates (with homogeneous effects):
                   $\mathcal{F}_{ql}(Y(i,j)) =
                   \mathcal{N}(\mu_{ql}(Y(i,j)),\sigma^2)$ where
                   $\mu_{ql}(Y(i,j))=\mu_{ql}+\beta^TY(i,j)$. This model uses
                   covariates and can model networks with real weight. The effect
                   of covariates is the same for all pair of classes.
          \subitem with covariates (with heterogeneous effects):
                   $\mathcal{F}_{ql}(Y(i,j)) =
                   \mathcal{N}(\mu_{ql}(Y(i,j)),\sigma^2)$ where
                   $\mu_{ql}(Y(i,j))=\mu_{ql}+\beta_{ql}^TY(i,j)$. This model
                   uses covariates and can model networks with real weight.
                   The effect of covariates is \emph{not} the same for all pair of
                   classes.
     \end{itemize}

  \subsection{Analysis of groups when covariates are used}

    Without covariates, groups are sets of nodes which have the same connectivity
    behavior (in probability), and groups can be easily interpretable using the
    connectivity matrix ($[\pi_{ql}]$, $[\lambda_{ql}]$ or $[\mu_{ql}]$).

    With covariates, groups are sets of nodes which have the same connectivity
    behavior (in probability) \emph{conditionally to covariates}. Two nodes of
    the same group can have different connectivity behavior due to different
    values of covariates.

    For a model with covariates, groups are covariate-residual groups. There are two
    points of view:

    \begin{itemize}

      \item
      the focus is on the effects of the covariates and groups model the (residual) connectivity which is \emph{not} explained by covariates,

      \item
      the focus is on the groups which helps in suggesting some sources of heterogeneity after correcting the artefact due to covariates.

    \end{itemize}

  One can test the effect of covariates using a likelihood ratio test between
  models with and without covariates.

\section{Estimation method}
  The estimation method is described in \citet{Mariada10}. The likelihood is not
  computable in a reasonable time, and a variational approximation is done and a
  \emph{variational expectation--maximization} is used. The ICL criterion is
  used for choosing  the number of groups, see \citet{Mariada10}.

    Some estimation implementation details which differ from the framework
    introduced in \citet{Mariada10} are explained here.

    \subsection{Initialization}
      As in the general case on \emph{expectation--maximization} algorithm, the initialization
      plays a major role in the quality of the local maximum found.

      In \citet{Mariada10}, the authors propose to use a hierarchical clustering
      to initialize the algorithm. In a real case of network analysis this
      initialization is often an extremal one (most of the initialized groups contain
      only one node) and the \emph{expectation--maximization} algorithm
      converges to a local maximum which may be far from the global maximum.

      The Absolute Value Spectral Clustering algorithm is consistent for finding
      groups in SBM (with Bernoulli probability law without covariates), see
      \citet{Rohe11}.
      We use the absolute spectral clustering to initialize the
      \emph{expectation--maximization} algorithm.

      When there are covariates, the spectral clustering is done on the residual
      graph, after eliminating the effect of covariates by regression.

    \subsection{Smoothing}

      To determine if an estimation for $Q$ groups has reached
       a bad local maximum, we use two findings:

      \begin{itemize}

        \item
        With an ascending number of groups, models are nested. A model
        with $Q$ groups can be interpreted as a model with $Q+1$ groups, so the
        likelihood must increase with Q.

        \item
        Empirical findings make us say that the ICL criterion is convex.

      \end{itemize}

      A reinitialization of the
      \emph{expectation--maximization} can be done. The new initialization is
      obtained in two ways:
      \begin{itemize}
        \item merging two groups of the $Q+1$ result (descend mode)
        \item splitting one group into two groups of the $Q-1$ result (ascend
        mode), this split is done by a spectral clustering of the residual graph
        on $Q-1$ groups.
      \end{itemize}

      There are two modes of reinitialization:
      \begin{itemize}
        \item the \code{minimal} one, reinitializations are done each time one of the
              two findings (see above) is not respected,
        \item the \code{exhaustive} one, all reinitializations are done; while it
              improves likelihood, this option is very time-consuming and cannot
              be used with non small graphs.
      \end{itemize}

    \subsection{Parallelism}
       Many steps of the estimation can be done
      independently:

      \begin{itemize}
        \item The \emph{expectation--maximization} algorithm for various $Q$
        \item Reinitialization in ascend and descend mode
      \end{itemize}

      Considering that computers and computing units have more than one logical
      processor, this implementation uses threads to parallelize the
      implementation as much as possible.

\section[wmixnet program]{\pkg{wmixnet} program}

    This section introduces the \pkg{wmixnet} program and the program usage.

    \subsection{Sources availability and installation}

    \pkg{wmixnet} is provided on the GNU General Public Licence version 3,
    and \proglang{C++} sources are available on the \pkg{wmixnet} page:

    \begin{itemize}
    \item \url{http://www.agroparistech.fr/mia/productions:logiciels}
    \item \url{http://www.agroparistech.fr/mia/productions:logiciel:wmixnet}
    \end{itemize}

    \pkg{wmixnet} should be installable from sources on any Linux
    distribution, when dependencies are provided:
    \begin{itemize}
      \item \pkg{IT++} library, used for matrix calculation. This library
      uses \pkg{blas} and \pkg{lapack}, well-known algebra libraries.
      \item \pkg{boost} library, for many aspects including parallelism.
    \end{itemize}

  \subsection{Input format}
    The input format is a plain text with the following specifications:
    \begin{itemize}
      \item each line describes a node
      \item for each line the first two columns describe the indexes of starting
            and ending nodes
      \item for each line the third column describes the weight of the edge
      \item for each line the fourth to end columns (if present) describe the
            covariates associated to the edge.
    \end{itemize}

    There are some constraints:
    \begin{itemize}
      \item node indexes must start from $1$ to the number of nodes
      \item each edge must have the same number of covariates.
    \end{itemize}

    If an edge is not present, and if no covariates are used, the corresponding
    lines can be omitted; otherwise the line must be present with a weight of zero.

    Functions are provided to write a file following these specifications,
    with adjacency matrices, and covariate matrices, for \pkg{GNU R}, and \pkg{MATLAB} or
    \pkg{GNU Octave}.

  \subsection{Output format}

    The output format contains the model parameters for all explored numbers of
    groups.

    Model parameters are:
    \begin{itemize}
      \item $\alpha$, the parameters of the multinomial distribution
      \item $\theta$, the parameters of the probability law of the edge weight
      conditionally to groups of nodes,
        \subitem for the Bernoulli model, $\theta=(\pi)$,
        \subitem for the Poisson model, $\theta=(\lambda)$,
        \subitem for the Poisson model with covariates $\theta=(\lambda,\beta)$,
        \subitem for the Gaussian model $\theta=(\mu,\sigma^2)$,
        \subitem for the Gaussian model with covariates
                 $\theta=(\mu,\sigma^2,\beta)$.
    \end{itemize}

    The output contains variational parameter estimates ($\tau$) which give
    the nodes membership in groups.

    The output also contains values of criteria such as pseudo-likelihood and the ICL
    criterion.

    \smallskip
    There are three output formats provided:
    \begin{itemize}

      \item
      Plain text output format (named text), which is a human readable file.

      \item
      \pkg{GNU R} file output format, which is an \pkg{GNU R} loadable file.  Nevertheless this
      file can be easily read by a human.

      \item
      \pkg{MATLAB} or \pkg{GNU Octave} file output format, which is a \pkg{MATLAB} and \pkg{GNU Octave}
      loadable file.  Nevertheless this file can be easily read by a human.
    \end{itemize}

  \subsection{Command line usage}
    \pkg{wmixnet} is usable with command line, and the following arguments must be
    provided:

    \begin{itemize}
      \item \code{{-}{-}input} to specify the input file,
      \item \code{{-}{-}symmetric} to indicate if the graph is an undirected graph if
            applicable,
      \item \code{{-}{-}model} to specify the model in
        \subitem \code{bernoulli} for Bernoulli without covariate
        \subitem \code{BH} for Bernoulli with covariates (homogeneous effects)
        \subitem \code{BI} for Bernoulli with covariates (heterogeneous
        effects)
        \subitem \code{poisson} for Poisson without covariate
        \subitem \code{PRMH} for Poisson with covariates (homogeneous effects)
        \subitem \code{PRMI} for Poisson with covariates (heterogeneous effects)
        \subitem \code{gaussian} for Gaussian without covariate
        \subitem \code{GRMH} for Gaussian with covariates (homogeneous
        effects)
        \subitem \code{GRMI} for Gaussian with covariates (heterogeneous
        effects)
      \item \code{{-}{-}Qmax} to specify the maximum number of groups, or
            \code{{-}{-}Qauto} to let the program choose the maximum number of
            groups,
      \item \code{{-}{-}smoothing} to specify the smoothing mode
        \subitem \code{none} no reinitialization is done (by default)
        \subitem \code{minimal} reinitializations are done for detected
                 problems
        \subitem \code{exhaustive} all reinitialization are done (time-consuming option, only for small graphs)
      \item \code{{-}{-}output} to specify the output file,
      \item \code{{-}{-}output-format} to specify the output format
        \subitem \code{text} (by default)
        \subitem \code{R} for \pkg{GNU R} loadable file
        \subitem \code{matlab} or \code{octave} which are synonymous for \pkg{MATLAB} and \pkg{GNU Octave} loadable file.
    \end{itemize}

  \subsection{Empirical complexity}

  Some simulations suggest the following estimation of complexity:

  \[
    t = C_{\textrm{model}}\,n^{2.46}\,g^{2.1}\,1.03^p
  \]

  with
  \begin{itemize}
    \item $t$ the total processor time (equivalent time on a mono-core computer,
    without parallelization, which executes only this job)
    \item $C_{\textrm{model}}$ a constant which depends on the model. Since
          absolute values are not pertinent, ratios are given:
      \subitem $\frac{C_{\textrm{poisson}}}{C_{\textrm{bernoulli}}}=3.9$
      \subitem $\frac{C_{\textrm{PRMH}}}{C_{\textrm{bernoulli}}}=21$
      \subitem $\frac{C_{\textrm{gaussian}}}{C_{\textrm{bernoulli}}}=840$
      \subitem $\frac{C_{\textrm{GRMH}}}{C_{\textrm{bernoulli}}}=1350$

          This ratio is dependent on the way each model is implemented. Some
          models allow us to vectorize some steps, have explicit maxima, and
          thus are significantly faster
    \item $n$ the number of nodes
    \item $g$ the number of groups found
    \item $p$ the number of covariates (the size of the covariate vector)
  \end{itemize}

  \subsection{Capacity of extension}

  In the \pkg{wmixnet} program, the estimation
  procedure and other model-common parts are implemented once. Only
  model-specific functions are present for each model. Therefore it is relatively easy
  to add other models in the \pkg{wmixnet} program.

\section{Example}

  Here we introduce the analysis of two ecological networks, already analyzed by
  \citet{Mariada10}.

  \subsection{The networks}

    The networks are two undirected, valued networks having parasitic fungal
    species ($n = 154$) and tree species ($n = 51$) as nodes, respectively.
    Edge strengths was defined as the number of shared host species and the
    number of shared parasitic species, respectively (see \citealp{Mariada10}
    for details).

  \subsection{Covariate data}

  For the tree species network, we know the taxonomic distance between tree
  species and the degree of geographic overlap between tree species
  distribution. For the fungal species network, we know the taxonomic distance
  between fungal species (see \citealp{Mariada10} for details).

  \subsection{Example of command line}

    For the analysis of the tree species network, for the Poisson model without
    covariates, the command line is:

    \begin{verbatim}
          wmixnet --input Trees.spm --symmetric \
            --model poisson \
            --Qauto --smoothing exhaustive \
            --output Trees.m --output-format octave
    \end{verbatim}

  \subsection{Results}

    \subsubsection{On the tree species network}

\begin{figure}[ht!]
        \begin{center}
          \def\tikscale{.85}\begin{tikzpicture}[gnuplot,every node/.style={scale=\tikscale},scale=\tikscale,]
\gpcolor{gp lt color border}
\gpsetlinetype{gp lt border}
\gpsetlinewidth{0.50}
\draw[gp path] (2.731,4.309)--(2.982,4.309);
\draw[gp path] (15.860,4.309)--(15.609,4.309);
\gpcolor{\gprgb{0}{0}{0}}
\node[gp node right,font=\gpfontsize{12.00pt}{14.40pt}] at (2.547,4.309) {-3000};
\gpcolor{gp lt color border}
\draw[gp path] (2.731,5.976)--(2.982,5.976);
\draw[gp path] (15.860,5.976)--(15.609,5.976);
\gpcolor{\gprgb{0}{0}{0}}
\node[gp node right,font=\gpfontsize{12.00pt}{14.40pt}] at (2.547,5.976) {-2500};
\gpcolor{gp lt color border}
\draw[gp path] (2.731,7.643)--(2.982,7.643);
\draw[gp path] (15.860,7.643)--(15.609,7.643);
\gpcolor{\gprgb{0}{0}{0}}
\node[gp node right,font=\gpfontsize{12.00pt}{14.40pt}] at (2.547,7.643) {-2000};
\gpcolor{gp lt color border}
\draw[gp path] (2.731,9.309)--(2.982,9.309);
\draw[gp path] (15.860,9.309)--(15.609,9.309);
\gpcolor{\gprgb{0}{0}{0}}
\node[gp node right,font=\gpfontsize{12.00pt}{14.40pt}] at (2.547,9.309) {-1500};
\gpcolor{gp lt color border}
\draw[gp path] (2.731,10.976)--(2.982,10.976);
\draw[gp path] (15.860,10.976)--(15.609,10.976);
\gpcolor{\gprgb{0}{0}{0}}
\node[gp node right,font=\gpfontsize{12.00pt}{14.40pt}] at (2.547,10.976) {-1000};
\gpcolor{gp lt color border}
\draw[gp path] (2.731,4.309)--(2.731,4.560);
\draw[gp path] (2.731,10.976)--(2.731,10.725);
\gpcolor{\gprgb{0}{0}{0}}
\node[gp node center,font=\gpfontsize{12.00pt}{14.40pt}] at (2.731,4.001) {0};
\gpcolor{gp lt color border}
\draw[gp path] (4.919,4.309)--(4.919,4.560);
\draw[gp path] (4.919,10.976)--(4.919,10.725);
\gpcolor{\gprgb{0}{0}{0}}
\node[gp node center,font=\gpfontsize{12.00pt}{14.40pt}] at (4.919,4.001) {2};
\gpcolor{gp lt color border}
\draw[gp path] (7.107,4.309)--(7.107,4.560);
\draw[gp path] (7.107,10.976)--(7.107,10.725);
\gpcolor{\gprgb{0}{0}{0}}
\node[gp node center,font=\gpfontsize{12.00pt}{14.40pt}] at (7.107,4.001) {4};
\gpcolor{gp lt color border}
\draw[gp path] (9.296,4.309)--(9.296,4.560);
\draw[gp path] (9.296,10.976)--(9.296,10.725);
\gpcolor{\gprgb{0}{0}{0}}
\node[gp node center,font=\gpfontsize{12.00pt}{14.40pt}] at (9.296,4.001) {6};
\gpcolor{gp lt color border}
\draw[gp path] (11.484,4.309)--(11.484,4.560);
\draw[gp path] (11.484,10.976)--(11.484,10.725);
\gpcolor{\gprgb{0}{0}{0}}
\node[gp node center,font=\gpfontsize{12.00pt}{14.40pt}] at (11.484,4.001) {8};
\gpcolor{gp lt color border}
\draw[gp path] (13.672,4.309)--(13.672,4.560);
\draw[gp path] (13.672,10.976)--(13.672,10.725);
\gpcolor{\gprgb{0}{0}{0}}
\node[gp node center,font=\gpfontsize{12.00pt}{14.40pt}] at (13.672,4.001) {10};
\gpcolor{gp lt color border}
\draw[gp path] (15.860,4.309)--(15.860,4.560);
\draw[gp path] (15.860,10.976)--(15.860,10.725);
\gpcolor{\gprgb{0}{0}{0}}
\node[gp node center,font=\gpfontsize{12.00pt}{14.40pt}] at (15.860,4.001) {12};
\gpcolor{gp lt color border}
\draw[gp path] (2.731,10.976)--(2.731,4.309)--(15.860,4.309)--(15.860,10.976)--cycle;
\gpcolor{\gprgb{0}{0}{0}}
\node[gp node center,rotate=90,font=\gpfontsize{12.00pt}{14.40pt}] at (1.289,7.642) {ICL criterion};
\node[gp node center,font=\gpfontsize{12.00pt}{14.40pt}] at (9.295,3.539) {number of groups};
\gpcolor{gp lt color border}
\node[gp node right,font=\gpfontsize{12.00pt}{14.40pt}] at (14.392,10.642) {Poisson};
\gpcolor{\gprgb{0}{0}{0}}
\gpsetlinetype{gp lt plot 0}
\draw[gp path] (14.576,10.642)--(15.492,10.642);
\draw[gp path] (3.825,4.721)--(4.919,7.957)--(6.013,8.450)--(7.107,8.888)--(8.201,9.087)%
  --(9.296,9.174)--(10.390,9.194)--(11.484,9.181)--(12.578,9.146)--(13.672,9.084)--(14.766,8.976);
\gpsetpointsize{4.00}
\gppoint{gp mark 1}{(3.825,4.721)}
\gppoint{gp mark 1}{(4.919,7.957)}
\gppoint{gp mark 1}{(6.013,8.450)}
\gppoint{gp mark 1}{(7.107,8.888)}
\gppoint{gp mark 1}{(8.201,9.087)}
\gppoint{gp mark 1}{(9.296,9.174)}
\gppoint{gp mark 1}{(10.390,9.194)}
\gppoint{gp mark 1}{(11.484,9.181)}
\gppoint{gp mark 1}{(12.578,9.146)}
\gppoint{gp mark 1}{(13.672,9.084)}
\gppoint{gp mark 1}{(14.766,8.976)}
\gppoint{gp mark 1}{(15.034,10.642)}
\gpcolor{gp lt color border}
\node[gp node right,font=\gpfontsize{12.00pt}{14.40pt}] at (14.392,10.334) {PRMH with taxo cov};
\gpcolor{\gprgb{1000}{0}{0}}
\gpsetlinetype{gp lt plot 1}
\draw[gp path] (14.576,10.334)--(15.492,10.334);
\draw[gp path] (3.825,6.366)--(4.919,9.089)--(6.013,9.482)--(7.107,9.662)--(8.201,9.620)%
  --(9.296,9.570)--(10.390,9.505)--(11.484,9.425)--(12.578,9.337)--(13.672,9.235)--(14.766,9.112);
\gppoint{gp mark 1}{(3.825,6.366)}
\gppoint{gp mark 1}{(4.919,9.089)}
\gppoint{gp mark 1}{(6.013,9.482)}
\gppoint{gp mark 1}{(7.107,9.662)}
\gppoint{gp mark 1}{(8.201,9.620)}
\gppoint{gp mark 1}{(9.296,9.570)}
\gppoint{gp mark 1}{(10.390,9.505)}
\gppoint{gp mark 1}{(11.484,9.425)}
\gppoint{gp mark 1}{(12.578,9.337)}
\gppoint{gp mark 1}{(13.672,9.235)}
\gppoint{gp mark 1}{(14.766,9.112)}
\gppoint{gp mark 1}{(15.034,10.334)}
\gpcolor{gp lt color border}
\node[gp node right,font=\gpfontsize{12.00pt}{14.40pt}] at (14.392,10.026) {PRMH with geo cov};
\gpcolor{\gprgb{0}{0}{1000}}
\gpsetlinetype{gp lt plot 2}
\draw[gp path] (14.576,10.026)--(15.492,10.026);
\draw[gp path] (3.825,5.212)--(4.919,8.065)--(6.013,8.643)--(7.107,9.096)--(8.201,9.211)%
  --(9.296,9.240)--(10.390,9.227)--(11.484,9.186)--(12.578,9.150)--(13.672,9.075)--(14.766,8.972);
\gppoint{gp mark 1}{(3.825,5.212)}
\gppoint{gp mark 1}{(4.919,8.065)}
\gppoint{gp mark 1}{(6.013,8.643)}
\gppoint{gp mark 1}{(7.107,9.096)}
\gppoint{gp mark 1}{(8.201,9.211)}
\gppoint{gp mark 1}{(9.296,9.240)}
\gppoint{gp mark 1}{(10.390,9.227)}
\gppoint{gp mark 1}{(11.484,9.186)}
\gppoint{gp mark 1}{(12.578,9.150)}
\gppoint{gp mark 1}{(13.672,9.075)}
\gppoint{gp mark 1}{(14.766,8.972)}
\gppoint{gp mark 1}{(15.034,10.026)}
\gpcolor{gp lt color border}
\node[gp node right,font=\gpfontsize{12.00pt}{14.40pt}] at (14.392,9.718) {PRMH with taxo and geo cov};
\gpcolor{\gprgb{0}{1000}{0}}
\gpsetlinetype{gp lt plot 3}
\draw[gp path] (14.576,9.718)--(15.492,9.718);
\draw[gp path] (3.825,6.871)--(4.919,9.137)--(6.013,9.543)--(7.107,9.662)--(8.201,9.613)%
  --(9.296,9.557)--(10.390,9.501)--(11.484,9.429)--(12.578,9.337)--(13.672,9.228)--(14.766,9.102);
\gppoint{gp mark 1}{(3.825,6.871)}
\gppoint{gp mark 1}{(4.919,9.137)}
\gppoint{gp mark 1}{(6.013,9.543)}
\gppoint{gp mark 1}{(7.107,9.662)}
\gppoint{gp mark 1}{(8.201,9.613)}
\gppoint{gp mark 1}{(9.296,9.557)}
\gppoint{gp mark 1}{(10.390,9.501)}
\gppoint{gp mark 1}{(11.484,9.429)}
\gppoint{gp mark 1}{(12.578,9.337)}
\gppoint{gp mark 1}{(13.672,9.228)}
\gppoint{gp mark 1}{(14.766,9.102)}
\gppoint{gp mark 1}{(15.034,9.718)}
\gpdefrectangularnode{gp plot 1}{\pgfpoint{2.731cm}{4.309cm}}{\pgfpoint{15.860cm}{10.976cm}}
\end{tikzpicture}

          \caption{ICL criterion values obtained for Poisson and Poisson with
                   covariates on the trees network}
          \label{f:treesICL}
        \end{center}
      \end{figure}
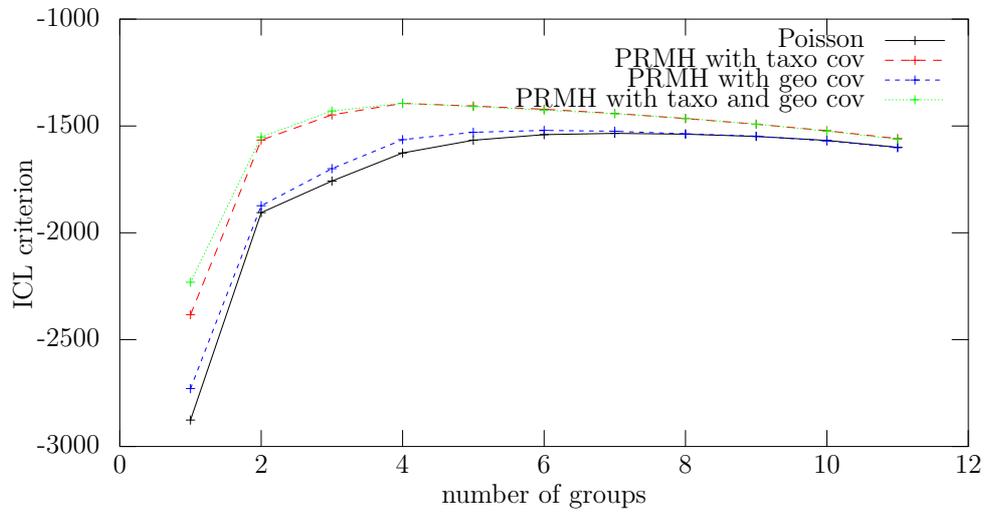

      In Figure~\ref{f:treesICL}, we plot the ICL criterion for Poisson model
      without and with covariates (taxonomic distance, geographical distance or
      both). For the model without covariates the maximum is reached with 7
      groups. With the geographical covariates, the maximum is reached for 6
      groups, with a little improvement of the ICL criterion. For the model with
      the taxonomic covariates, the maximum is reached for 4 groups, with a
      larger improvement of the ICL criterion. Adding the geographical covariates to
      the taxonomic covariates does not improve the criterion.

      Ecological interpretations are presented in \citet{Mariada10}.

      \subsubsection{On the fungi network}

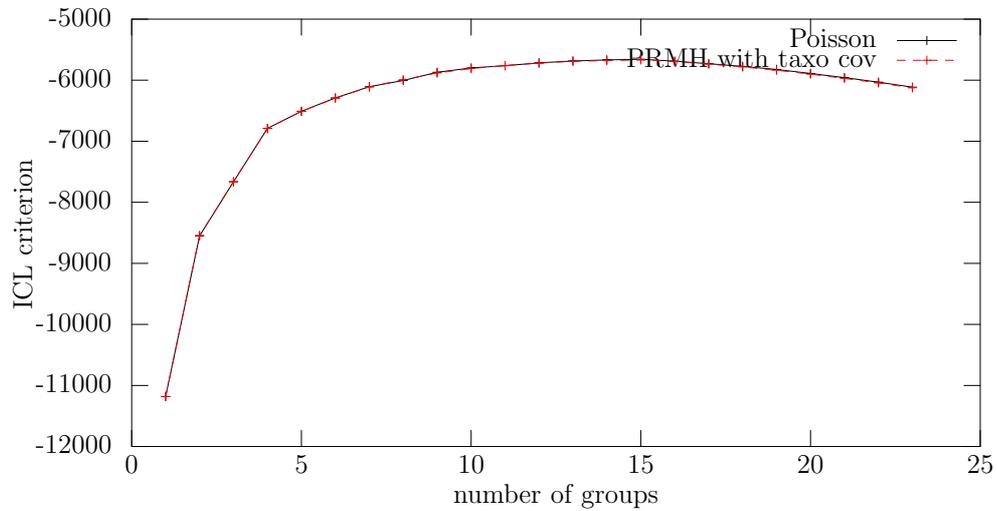
\begin{figure}[ht!]
        \begin{center}
          \def\tikscale{.85}\begin{tikzpicture}[gnuplot,every node/.style={scale=\tikscale},scale=\tikscale,]
\gpcolor{gp lt color border}
\gpsetlinetype{gp lt border}
\gpsetlinewidth{0.50}
\draw[gp path] (2.731,4.309)--(2.982,4.309);
\draw[gp path] (15.860,4.309)--(15.609,4.309);
\gpcolor{\gprgb{0}{0}{0}}
\node[gp node right,font=\gpfontsize{12.00pt}{14.40pt}] at (2.547,4.309) {-12000};
\gpcolor{gp lt color border}
\draw[gp path] (2.731,5.261)--(2.982,5.261);
\draw[gp path] (15.860,5.261)--(15.609,5.261);
\gpcolor{\gprgb{0}{0}{0}}
\node[gp node right,font=\gpfontsize{12.00pt}{14.40pt}] at (2.547,5.261) {-11000};
\gpcolor{gp lt color border}
\draw[gp path] (2.731,6.214)--(2.982,6.214);
\draw[gp path] (15.860,6.214)--(15.609,6.214);
\gpcolor{\gprgb{0}{0}{0}}
\node[gp node right,font=\gpfontsize{12.00pt}{14.40pt}] at (2.547,6.214) {-10000};
\gpcolor{gp lt color border}
\draw[gp path] (2.731,7.166)--(2.982,7.166);
\draw[gp path] (15.860,7.166)--(15.609,7.166);
\gpcolor{\gprgb{0}{0}{0}}
\node[gp node right,font=\gpfontsize{12.00pt}{14.40pt}] at (2.547,7.166) {-9000};
\gpcolor{gp lt color border}
\draw[gp path] (2.731,8.119)--(2.982,8.119);
\draw[gp path] (15.860,8.119)--(15.609,8.119);
\gpcolor{\gprgb{0}{0}{0}}
\node[gp node right,font=\gpfontsize{12.00pt}{14.40pt}] at (2.547,8.119) {-8000};
\gpcolor{gp lt color border}
\draw[gp path] (2.731,9.071)--(2.982,9.071);
\draw[gp path] (15.860,9.071)--(15.609,9.071);
\gpcolor{\gprgb{0}{0}{0}}
\node[gp node right,font=\gpfontsize{12.00pt}{14.40pt}] at (2.547,9.071) {-7000};
\gpcolor{gp lt color border}
\draw[gp path] (2.731,10.024)--(2.982,10.024);
\draw[gp path] (15.860,10.024)--(15.609,10.024);
\gpcolor{\gprgb{0}{0}{0}}
\node[gp node right,font=\gpfontsize{12.00pt}{14.40pt}] at (2.547,10.024) {-6000};
\gpcolor{gp lt color border}
\draw[gp path] (2.731,10.976)--(2.982,10.976);
\draw[gp path] (15.860,10.976)--(15.609,10.976);
\gpcolor{\gprgb{0}{0}{0}}
\node[gp node right,font=\gpfontsize{12.00pt}{14.40pt}] at (2.547,10.976) {-5000};
\gpcolor{gp lt color border}
\draw[gp path] (2.731,4.309)--(2.731,4.560);
\draw[gp path] (2.731,10.976)--(2.731,10.725);
\gpcolor{\gprgb{0}{0}{0}}
\node[gp node center,font=\gpfontsize{12.00pt}{14.40pt}] at (2.731,4.001) {0};
\gpcolor{gp lt color border}
\draw[gp path] (5.357,4.309)--(5.357,4.560);
\draw[gp path] (5.357,10.976)--(5.357,10.725);
\gpcolor{\gprgb{0}{0}{0}}
\node[gp node center,font=\gpfontsize{12.00pt}{14.40pt}] at (5.357,4.001) {5};
\gpcolor{gp lt color border}
\draw[gp path] (7.983,4.309)--(7.983,4.560);
\draw[gp path] (7.983,10.976)--(7.983,10.725);
\gpcolor{\gprgb{0}{0}{0}}
\node[gp node center,font=\gpfontsize{12.00pt}{14.40pt}] at (7.983,4.001) {10};
\gpcolor{gp lt color border}
\draw[gp path] (10.608,4.309)--(10.608,4.560);
\draw[gp path] (10.608,10.976)--(10.608,10.725);
\gpcolor{\gprgb{0}{0}{0}}
\node[gp node center,font=\gpfontsize{12.00pt}{14.40pt}] at (10.608,4.001) {15};
\gpcolor{gp lt color border}
\draw[gp path] (13.234,4.309)--(13.234,4.560);
\draw[gp path] (13.234,10.976)--(13.234,10.725);
\gpcolor{\gprgb{0}{0}{0}}
\node[gp node center,font=\gpfontsize{12.00pt}{14.40pt}] at (13.234,4.001) {20};
\gpcolor{gp lt color border}
\draw[gp path] (15.860,4.309)--(15.860,4.560);
\draw[gp path] (15.860,10.976)--(15.860,10.725);
\gpcolor{\gprgb{0}{0}{0}}
\node[gp node center,font=\gpfontsize{12.00pt}{14.40pt}] at (15.860,4.001) {25};
\gpcolor{gp lt color border}
\draw[gp path] (2.731,10.976)--(2.731,4.309)--(15.860,4.309)--(15.860,10.976)--cycle;
\gpcolor{\gprgb{0}{0}{0}}
\node[gp node center,rotate=90,font=\gpfontsize{12.00pt}{14.40pt}] at (1.105,7.642) {ICL criterion};
\node[gp node center,font=\gpfontsize{12.00pt}{14.40pt}] at (9.295,3.539) {number of groups};
\gpcolor{gp lt color border}
\node[gp node right,font=\gpfontsize{12.00pt}{14.40pt}] at (14.392,10.642) {Poisson};
\gpcolor{\gprgb{0}{0}{0}}
\gpsetlinetype{gp lt plot 0}
\draw[gp path] (14.576,10.642)--(15.492,10.642);
\draw[gp path] (3.256,5.089)--(3.781,7.594)--(4.306,8.438)--(4.832,9.272)--(5.357,9.539)%
  --(5.882,9.747)--(6.407,9.922)--(6.932,10.019)--(7.457,10.147)--(7.983,10.215)--(8.508,10.250)%
  --(9.033,10.294)--(9.558,10.323)--(10.083,10.340)--(10.608,10.350)--(11.134,10.320)--(11.659,10.281)%
  --(12.184,10.239)--(12.709,10.191)--(13.234,10.129)--(13.759,10.064)--(14.285,9.993)--(14.810,9.915);
\gpsetpointsize{4.00}
\gppoint{gp mark 1}{(3.256,5.089)}
\gppoint{gp mark 1}{(3.781,7.594)}
\gppoint{gp mark 1}{(4.306,8.438)}
\gppoint{gp mark 1}{(4.832,9.272)}
\gppoint{gp mark 1}{(5.357,9.539)}
\gppoint{gp mark 1}{(5.882,9.747)}
\gppoint{gp mark 1}{(6.407,9.922)}
\gppoint{gp mark 1}{(6.932,10.019)}
\gppoint{gp mark 1}{(7.457,10.147)}
\gppoint{gp mark 1}{(7.983,10.215)}
\gppoint{gp mark 1}{(8.508,10.250)}
\gppoint{gp mark 1}{(9.033,10.294)}
\gppoint{gp mark 1}{(9.558,10.323)}
\gppoint{gp mark 1}{(10.083,10.340)}
\gppoint{gp mark 1}{(10.608,10.350)}
\gppoint{gp mark 1}{(11.134,10.320)}
\gppoint{gp mark 1}{(11.659,10.281)}
\gppoint{gp mark 1}{(12.184,10.239)}
\gppoint{gp mark 1}{(12.709,10.191)}
\gppoint{gp mark 1}{(13.234,10.129)}
\gppoint{gp mark 1}{(13.759,10.064)}
\gppoint{gp mark 1}{(14.285,9.993)}
\gppoint{gp mark 1}{(14.810,9.915)}
\gppoint{gp mark 1}{(15.034,10.642)}
\gpcolor{gp lt color border}
\node[gp node right,font=\gpfontsize{12.00pt}{14.40pt}] at (14.392,10.334) {PRMH with taxo cov};
\gpcolor{\gprgb{1000}{0}{0}}
\gpsetlinetype{gp lt plot 1}
\draw[gp path] (14.576,10.334)--(15.492,10.334);
\draw[gp path] (3.256,5.087)--(3.781,7.601)--(4.306,8.441)--(4.832,9.270)--(5.357,9.536)%
  --(5.882,9.743)--(6.407,9.918)--(6.932,10.031)--(7.457,10.137)--(7.983,10.204)--(8.508,10.253)%
  --(9.033,10.289)--(9.558,10.319)--(10.083,10.329)--(10.608,10.341)--(11.134,10.314)--(11.659,10.275)%
  --(12.184,10.231)--(12.709,10.179)--(13.234,10.118)--(13.759,10.052)--(14.285,9.981)--(14.810,9.905);
\gppoint{gp mark 1}{(3.256,5.087)}
\gppoint{gp mark 1}{(3.781,7.601)}
\gppoint{gp mark 1}{(4.306,8.441)}
\gppoint{gp mark 1}{(4.832,9.270)}
\gppoint{gp mark 1}{(5.357,9.536)}
\gppoint{gp mark 1}{(5.882,9.743)}
\gppoint{gp mark 1}{(6.407,9.918)}
\gppoint{gp mark 1}{(6.932,10.031)}
\gppoint{gp mark 1}{(7.457,10.137)}
\gppoint{gp mark 1}{(7.983,10.204)}
\gppoint{gp mark 1}{(8.508,10.253)}
\gppoint{gp mark 1}{(9.033,10.289)}
\gppoint{gp mark 1}{(9.558,10.319)}
\gppoint{gp mark 1}{(10.083,10.329)}
\gppoint{gp mark 1}{(10.608,10.341)}
\gppoint{gp mark 1}{(11.134,10.314)}
\gppoint{gp mark 1}{(11.659,10.275)}
\gppoint{gp mark 1}{(12.184,10.231)}
\gppoint{gp mark 1}{(12.709,10.179)}
\gppoint{gp mark 1}{(13.234,10.118)}
\gppoint{gp mark 1}{(13.759,10.052)}
\gppoint{gp mark 1}{(14.285,9.981)}
\gppoint{gp mark 1}{(14.810,9.905)}
\gppoint{gp mark 1}{(15.034,10.334)}
\gpdefrectangularnode{gp plot 1}{\pgfpoint{2.731cm}{4.309cm}}{\pgfpoint{15.860cm}{10.976cm}}
\end{tikzpicture}

          \caption{ICL criterion values obtained for the Poisson model and the Poisson with
                   covariates model on the fungus species network.}
          \label{f:fungiICL}
        \end{center}
      \end{figure}

      In Figure~\ref{f:fungiICL}, we plot the ICL criterion for the Poisson model
      without and with covariates (taxonomic distance). For the model with and
      without taxonomic covariates the maximum is reached with 15 groups in both
      cases. There is no real improvement by adding the taxonomic covariates to
      the model.

      Ecological interpretations are presented in \citet{Mariada10}.

\section*{Acknowledgments}
We thank Dominique Piou, Corinne Vacher, and the Département Santé des Forêts
of the French Ministère de l'Agriculture et de la Pêche for allowing us to
use the network data.

We thank Stephane Robin for providing explanation and advice on the estimation
procedure and Pierre Barbillon for the Bernoulli model with covariates (BH and
BI) and for helping find some bugs in the estimation procedure.

\clearpage
  \bibliographystyle{abbrvnat}
  \bibliography{biblio}

\end{document}